\documentclass{nature-2}

\pdfoutput=1

\usepackage{graphicx}
\usepackage{verbatim}
\usepackage{subfigure}
\usepackage{tikz}
\usepackage{amsmath}
\usepackage{amssymb}
\usepackage{array}
\usepackage{tabularx}
\usepackage{multirow}
\usepackage{multicol}
\usepackage{multibox}
\usepackage{rotating}
\usepackage{color}
\usepackage{layout}
\usepackage{epstopdf}
\usepackage{afterpage}

\title{Predicting Whole Forest Structure, Primary Productivity, and Biomass Density From Maximum Tree Size and Resource Limitations}

\author{Christopher P. Kempes$^{1,2}$, Sungho Choi$^3$, William Dooris $^1$ \& Geoffrey B. West$^{4,5}$}

\begin{document}

\maketitle

\begin{affiliations}
 \item Control and Dynamical Systems, California Institute of Technology, Pasadena, CA 91125, USA
 \item NASA Ames Research Center, Moffett Field, CA 94035, USA
 \item Department of Earth and Environment, Boston University, Boston, MA 02215, USA
 \item Santa Fe Institute, Santa Fe, New Mexico 87501, USA
 \item Department of Mathematics, Imperial College, London, UK
\end{affiliations}

\begin{abstract}
In the face of uncertain biological response to climate change and the many critiques concerning model complexity it is increasingly important to develop predictive mechanistic frameworks that capture the dominant features of ecological communities and their dependencies on environmental factors. This is particularly important for critical global processes such as biomass changes, carbon export, and biogenic climate feedback \cite{schimel,chapin,allen,bonan2008,nemani,charney,fisher,medvigy,cox}. Past efforts have successfully understood a broad spectrum of plant and community traits across a range of biological diversity and body size, including tree size distributions and maximum tree height, from mechanical, hydrodynamic, and resource constraints \cite{brown,mech-stab-2,vp,plantcom-1,plantcom-2,kempes}. Recently it was shown that global scaling relationships for net primary productivity are correlated with local meteorology and the overall biomass density within a forest \cite{michaletz}. Along with previous efforts, this highlights the connection between widely observed allometric relationships and predictive ecology. An emerging goal of ecological theory is to gain maximum predictive power with the least number of parameters \cite{marquet}. Here we show that the explicit dependence of such critical quantities can be systematically predicted knowing just the size of the largest tree. This is supported by data showing that forests converge to our predictions as they mature. Since maximum tree size can be calculated from local meteorology \cite{kempes} this provides a general framework for predicting the generic structure of forests from local environmental parameters thereby addressing a range of critical Earth-system questions.
\end{abstract}

Terrestrial net primary productivity (NPP) is one of the key processes affecting the overall climate system, and given its dependence on local meteorology, defines one of the most important global feedback processes \cite{schimel,chapin,allen,bonan2008,nemani,fisher,medvigy}. Classically it has been shown that terrestrial biomass production correlates with precipitation and temperature \cite{schuur,huxman,ponce,berry}, but more recent efforts have pointed to existing biomass density as the strongest determinant of NPP and highlight the indirect consequences of the local environment potentially complicating the prediction of the biological response to climate change \cite{michaletz}. Here we show that biomass density and net primary productivity can both be determined from maximum tree size which in turn has been shown to be a calculable complex function of local climate \cite{kempes} and consequently a strong predictor of biomass density, as supported by empirical evidence \cite{stegen}. Through these relationships we also predict the observed relationship between NPP and biomass density, where these two features taken together represent the dominant climate feedbacks either through carbon uptake or the biological effects on atmospheric water and heat fluxes \cite{charney}. 

In our derivations we first express all scaling relationships in terms of general exponents without specifying their value. Although many of these can themselves be derived from an underlying theoretical framework, we use both theoretical as well as empirical values for these exponents in making predictions. These are compared with data spanning the entire continental United States representing a large number of whole forests subdivided into over $9\times10^{5}$ one km$^{2}$ size grid cells covering all 36 ecoregions \cite{cleland}. We discuss how these predictions depend on differences in scaling relationships for a given species or environment and show how forest maturity affects the observed scaling results. 

Recent work has shown that the general spacing of trees can be derived by combining mortality processes and the geometric scaling of vascular plants with the idea that the resource use per unit area is approximately constant in a given forest \cite{plantcom-1,plantcom-2}. Consequently, the average distance, $d_{k}$, between individual trees within a given size class is predicted to scale with trunk radius:
\begin{equation}
d_{k}=c_{1}r_{k}^{\alpha_{d}} 
\label{spacing}
\end{equation}
where $k$ denotes a linearly binned size class of trees\cite{plantcom-1,plantcom-2}, $c_{1}$ is a constant, and $r_{k}$ is the trunk radius, which is related to tree height, $h_{k}$, and the mass of the tree, $m_{k}$, by $r_{k}=c_{h}h_{k}^{\alpha_{h}}=c_{m}m_{k}^{\alpha_{m}}$ where $c_{h}$ and $c_{m}$ are also constants, and $\alpha_{h}$ and $\alpha_{m}$ are scaling exponents \cite{vp,kempes}. In this relationship $\alpha_{d}$ is the scaling exponent which, from theoretical and empirical work, is expected to be close to $1$ \cite{plantcom-1,plantcom-2}. Its possible variation could depend on a variety of factors from the scaling of overall metabolic rate to that of the canopy.  Since equation \ref{spacing} is based on the idea that resource consumption per unit area is constant this further implies that $c_{1}$ should depend on resource availability. This can be shown by noting that the largest tree in the forest completely dominates its area and does not share resources with any neighboring tree (if it did then it would be possible for a larger tree to exist) \cite{kempes}. In a similar fashion, the radial extent of the roots of the largest tree, $r_{root,max}$, must also define the separation of trees, in which case we have for the largest tree
\begin{equation}
d_{max}\equiv r_{root,max}=c_{1}r_{max}^{\alpha_{d}}
\end{equation}  
leading to $c_{1}=\frac{r_{root,max}}{r_{max}^{\alpha_{d}}}$. In previous work it was shown that the root radius is related to the trunk radius, $r_{root}^{max}=c_{root}r_{max}^{\alpha_{r}}$, which in turn can be determined by the local resource environment \cite{kempes}. Previous work has shown that the scaling is given by $\alpha_{r}=2/3$ and $c_{root}=\beta_{3}^{1/4}l_{N}r_{N}^{-2/3}/(1-n^{-1/3})$ where the quantities $\beta_{3}$, $l_{N}$, $r_{N}$, and $n$ are all basic normalization constants of scaling whose precise definition can be found in Ref. \cite{kempes}. Through its dependence on $r_{max}$ , $c_{1}$ is a complicated but calculable function of solar radiation, temperature, humidity, precipitation, and altitude. We can test this result using the distributions in Ref. \cite{plantcom-1} where the largest tree has a trunk radius of $54.5$ cm. Using $\alpha_{r}=2/3$, $\alpha_{d}=1$ and the above expression for $c_{root}$, we find $c_{1}=73$, in excellent agreement with the corresponding published best fit range of $c_{1}=63$ to $78$ from Ref. \cite{plantcom-1}. These results illustrate that the absolute distance between trees within any given size class can be determined solely from the size of the largest tree.

Previous work \cite{plantcom-1} has also shown that the number of trees of a given size scales with trunk radius as
\begin{equation}
n_{k}=c_{5}r_{k}^{\alpha_{n}}.
\end{equation} 
where, by definition, there is only a single largest tree (so $n_{max}=1$); $c_{5}$ is a constant given by $c_{5}=r_{max}^{-\alpha_{n}}$ with $\alpha_{n} = -2$ from previous work. This relationship explicitly shows that the overall properties of the entire forest are determined by the size of the largest individual. Because these results define the overall distribution of trees we can further show that numerous other features of the forest can be straightforwardly determined from the maximum size, notably the average biomass density and net primary productivity, both of which are of critical importance to atmospheric feedback processes and climate dynamics. We first show how whole forest totals depend on maximum size and from these we can calculate various macro-scale averages of the forest. 

The mass of an individual tree within a given size class is given by $m_{k}=\left(r_{k}/c_{m}\right)^{1/\alpha_{m}}$ so the total mass of all trees in the forest is 
\begin{equation}
M_{tot}=\int_{r_{0}}^{r_{max}}n_{k}m_{k}dr= \frac{r_{max}^{-\alpha_{n}}}{c_{m}^{1/\alpha_{m}}}\left(\frac{r_{max}^{1+\alpha_{n}+1/\alpha_{m}}-r_{0}^{1+\alpha_{n}+1/\alpha_{m}}}{1+\alpha_{n}+1/\alpha_{m}}\right)
\label{masstot}
\end{equation}
which means that for large maximum size the total forest mass scales as $M_{tot}\propto r_{max}^{1+1/\alpha_{m}}$ assuming that $1+\alpha_{n}+1/\alpha_{m}>0$ as supported by the existing exponent values discussed later. Given that the metabolic rate of an individual tree is given by $B_{k}=B_{0} r_{k}^{\alpha_{B}}$ the total metabolic rate of the entire forest is 
\begin{equation}
B_{tot}=\int_{r_{0}}^{r_{max}}n_{k}B_{k}dr =  B_{0}r_{max}^{-\alpha_{n}} \left(\frac{r_{max}^{1+\alpha_{n}+\alpha_{B}}-r_{0}^{1+\alpha_{n}+\alpha_{B}}}{1+\alpha_{n}+\alpha_{B}}\right)
\end{equation}
leading to $B_{tot}\propto r_{max}^{1+\alpha_{B}}$. Taken together with equation \ref{masstot} this implies that whole forest metabolism scales with whole forest mass with an exponent of $\frac{1+\alpha_{B}}{1+1/\alpha_{m}}$. A common expectation for the scaling of metabolism with step radius is $\alpha_{B}=2$ \cite{vp} (equivalent to scaling as $m_{k}^{3/4}$ \cite{vp,mori}), and it is also commonly assumed that $\alpha_{m}=3/8$ \cite{vp} in which case whole forest metabolism scales with whole forest mass following a power of $9/11$ which is larger than the single tree scaling of $3/4$, but still less than linear. 

The total area required to observe a full distribution of tree sizes can be found by first noting that the largest tree dominates its resource area and thus $A_{max}=\pi r_{root,max}^{2}$, where the root radius is related to maximum size following $r_{root,max}=c_{root}r_{max}^{\alpha_{r}}$ \cite{kempes}. Noting that space-filling implies that the total area of each size class, $A_{k}$, is the same for all $k$, as is the total metabolic rate, $\frac{n_{k}B_{k}}{A_{k}}=\frac{B_{max}}{A_{max}}$, requires that $A_{k}=A_{max}$. Combining these relationships gives
\begin{equation}
A_{tot}=\int_{r_{0}}^{r_{max}} A_{k} dr = \pi c_{root}^{2} r_{max}^{2\alpha_{r}}\left(r_{max}-r_{0}\right) .
\end{equation}
which scales as $A_{tot}\propto r_{max}^{1+2\alpha_{r}}$. 

These aggregate quantities allow us to calculate the average biomass density, $D$:
\begin{equation}
D= \frac{M_{tot}}{A_{tot}}\propto r_{max}^{1/\alpha_{m}-2\alpha_{r}}
\end{equation}
where it is convenient to define $\alpha_{D}\equiv1/\alpha_{m}-2\alpha_{r}$. This relationship depends on only two scaling exponents, that of the relationship between mass and trunk radius and that of root radius (the lateral extent of the roots) to trunk radius. 

In order to make predictions across forests we need to estimate and assess the exponents in this relationship which in general can depend on species, environment, forest age, and competitive effects (e.g. \cite{chave,coomes,lin,mori,niklas,mech-stab-2,plantcom-1,plantcom-2}).  Previous work has shown  that theoretically $\alpha_{m}\approx3/8$ \cite{vp,niklas} and $\alpha_{r}\approx2/3$ \cite{kempes,niklas,mech-stab-2,salas} consistent with empirical data; together these predict $D\propto r_{max}^{4/3}$. An alternative way of estimating $\alpha_{m}$ is to note that many studies report the metabolic scaling relationship with overall tree mass, $B\propto m^{\alpha_{1}}$, which, together with the relationships described above, gives $\alpha_{m}=\alpha_{1}/\alpha_{B}$.  Theoretical predictions give $\alpha_{B}=2$ \cite{vp, penergetics, savage} which is in agreement with measurements for a proxy of metabolism which give $\alpha_{B}=1.78$ in one study \cite{penergetics} and  $\alpha_{B}=1.77$ (with a confidence interval of $1.38$ to $2.16$) in a second study \cite{savage}. For the metabolic scaling exponent with mass, $\alpha_{1}$, there are a wide variety of studies and predictions with a common theoretical assessment being that $\alpha_{1}\approx3/4$ (e.g. Ref. \cite{vp}). Precise measurements of individual tree respiration for a variety of species and a diversity of ecosystems (tropical to boreal) have shown a tight relationship between body mass and metabolism where the exponent smoothly varied from roughly $1$ in small trees to $3/4$ in large trees, as predicted. Furthermore, given previous analyses of the relationship between tree density and average biomass \cite{penergetics} we estimate from Ref. \cite{lin} that across modes of plant interaction $\alpha_{1}$ ranges between roughly $2/3$ and $0.91$. Measurements in animals, which we provide as a reference, have found that the metabolic exponent varies as widely as $0.25$ to $1.25$ \cite{coomes}. These results imply that $\alpha_{m}$ could vary in various circumstances along with the resulting scaling for $D$. For example, taking $\alpha_{1}$ between $2/3$ and $0.91$ and $\alpha_{B}$ between $1.78$ and $2$ predicts a range for $\alpha_{m}$ of roughly $1/3$ to $1/2$ which spans the expected value of $3/8$ and agrees with measurements that show a convergence to $3/8$ as trees become large (with asymptotic behavior at small sizes) \cite{niklas}. Similarly, it has been shown that $m\propto \left(\rho r^{2} h\right)^{0.976}$ where $\rho$ is wood density \cite{chave}. This result would give $\alpha_{m}\approx3/8$ given the common assumption that $h\propto r^{2/3}$ (e.g. \cite{vp}) and also shows the importance of species-level differences which could affect wood density. Taken together with $\alpha_{r}$ these calculations predict that the biomass density scaling could vary between $D\propto r_{max}^{2/3}$ and $D\propto r_{max}^{5/3}$.

To test this result for the dependence of biomass density on maximum size we gathered data for the continental United States \cite{blackard,gray} spanning arid environments to old-growth temperate rain forests. Our derivations predict scaling relationships in whole forest properties and so each point in our database and plots represent paired values for entire forests where the predicted trends should manifest across forests from many different environments (i.e. many different biomass densities). The data represent a variety of forest ages which affects both the distribution of tree sizes \cite{plantcom-1,plantcom-2} and how well the observed maximum size represents the true maximum of a given environment. For example, previous work shows that forest age, often a measure of recent disturbance, can affect both the intercept \cite{michaletz} and exponent \cite{plantcom-1} of scaling relationships. Using stand age as a proxy for forest disturbance we should expect significant error between the predicted scaling relationships and observations for young forests suggesting that we examine the scaling of biomass density within forest age categories \cite{pan}. We find that as forests mature the observed scaling exponents between biomass density and maximum observed trunk radius converge to the theoretical prediction of $4/3$ with some variation that could be the result of species-, environment-, or forest-level differences in the underlying scaling exponents ({\bf Figure} \ref{scaling-exponents}{\bf \subref{biomass-scaling} and \ref{scaling-exponents}\subref{biomass-age}}, and {\bf Table 1} which provides information for each age bin and fit). These results illustrate the predictive power of the maximum tree size for overall forest structure and also the strong effect that forest maturity has on the deviations between observed and steady-state forest properties. In {\bf Figure \ref{scaling-exponents}\subref{biomass-age}} we have also shown other possibilities for $\alpha_{D}$ based on the above calculations.

Similarly, it has been shown that the net primary productivity of a forest scales with plant mass in roughly the same way that metabolism does \cite{mech-stab-2,enquist2007} and thus the two are linearly related: $NPP=c_{p}B$, where $c_{p}$ is a constant. The average NPP is then given by
\begin{equation}
NPP=c_{p}\frac{B_{tot}}{A_{tot}}\propto r_{max}^{\alpha_{B}-2\alpha_{r}}.
\end{equation}
where we define $\alpha_{NPP}\equiv\alpha_{B}-2\alpha_{r}$. The theoretical values discussed earlier for the exponents would predict that $NPP\propto r_{max}^{2/3}$. Again considering possible variation in the exponents, we find $NPP\propto r_{max}^{0.44}$ when we take $\alpha_{B}=1.77$.

Data \cite{zhao,gray} show that these predicted scaling relationships also hold for forests that reach a certain age of maturity ({\bf Figure} \ref{scaling-exponents}{\bf \subref{npp-scaling} and \ref{scaling-exponents}\subref{npp-age}}). Taken together our relationships for biomass density and primary productivity imply that they are related via
\begin{equation}
NPP\propto D^{\left(\alpha_{B}-2\alpha_{r}\right)/\left(1/\alpha_{m}-2\alpha_{r}\right)}
\end{equation}
which, using predicted exponents, would give $NPP\propto D^{1/2}$. This is  well supported by data from a range of forests analyzed here ({\bf Figure} \ref{scaling-exponents}{\bf\subref{npp-biomass-scaling} and \ref{scaling-exponents}\subref{npp-biomass-age}}) as well as in previously published data \cite{michaletz} ({\bf  Figure} \ref{enquist-npp}). 

It should be noted that all of the scaling relationships for whole forest averages are consistent with the biomass density and NPP of the maximum size tree, which dominates its resource area and should be representative of the overall forest. For example, for the largest tree $m_{max}/A_{max}\propto r_{max}^{1\alpha_{m}-2\alpha_{r}}$, which is consistent with the whole forest average biomass density. In the above derivations we have allowed each scaling relationship to have an independent exponent value for the greatest degree of generality. However, it should also be stressed that many of the predicted exponents are related to, or derivative of, other exponents via theoretical derivations and empirical results.

It is interesting to note that for both biomass density, D, and NPP the size-distribution scaling exponent, $\alpha_{n}$, is eliminated in the limit of large trees. For completeness we discuss here the values of $\alpha_{n}$ which might be needed for calculations of $M_{tot}$, $B_{tot}$, or $A_{tot}$. It has been shown empirically that the size distribution scaling ranges from $\alpha_{n}\approx-3.5$ to $\alpha_{n}\approx-2$  with young forests clustered around $-3$ and mature forests around $-2$ \cite{plantcom-1}. A detailed study of forests of different age showed a convergence to the theoretically predicted value of $-2$ as forests mature, similar to the perspectives presented in this study. These results along with our own illustrate that many scaling predictions may hold only in mature forests with low disturbance. Similarly, many other factors such as species differences may affect scaling relationships as discussed earlier and thus caution should be applied in the application of any scaling relationship to a particular context. For example, it has been shown that the plant density exponent can vary from $-1.96$ in angiosperm communities to $-0.78$ in conifer communities \cite{mech-stab-2}.

As already discussed, recent work has shown how tree allometry can be used to derive the energy and water budgets of a tree and  to predict maximum tree size as a function of local meteorological conditions \cite{kempes}. Because maximum tree size determines the overall density of biomass, growth, and metabolic rate within a forest, these previous results provide a simple allometric framework for predicting whole forest features from local resource constraints ({\bf Figure} \ref{environmental-predictions}{\bf \subref{framework-schematic}}). Notably $NPP$ and biomass density can be written as functions of only maximum size which is then a complicated, but calculable, function of environmental factors,
\begin{equation}
NPP\propto r_{max}^{\alpha_{B}-2\alpha_{r}}.
\end{equation}
\begin{equation}
D\propto r_{max}^{1/\alpha_{m}-2\alpha_{r}}.
\end{equation}
\begin{equation}
r_{max}=f\left(p,t,rh,s\right),
\end{equation}
illustrating that the previously published dependencies \cite{michaletz} can be greatly simplified ($p$ is precipitation, $t$ is temperature, $rh$ is relative humidity, and $s$ is solar radiation). To illustrate this point we used the model in Ref. \cite{kempes} to predict $r_{max}$ as a function of only precipitation (holding all other climatic variables to global averages) and from $r_{max}$ we are able to predict the dependence of NPP on precipitation, in good agreement with published data \cite{michaletz} and our own analysis of the United States ({\bf Figure} \ref{environmental-predictions}{\bf \subref{precip-scaling}}). We provide this analysis to illustrate how allometric frameworks can be used in the future to predict forest features from local climate. It is important to note that in these predictions it is necessary to consider the full combination of local climatic variables as increases in one feature, for example temperature, can have opposing effects in different climates because of differing impacts on the tree's energy and water budgets. This is illustrated by previous studies that show differing correlations between temperature and biomass density in different forests \cite{stegen,keith} and partially explains why Michaletz et al. find only strong correlations with precipitation. Indeed, future efforts are needed to understand how NPP and biomass density are related to climate combinations in connection with many past studies (e.g. \cite{schuur,huxman,ponce,berry, michaletz,liu,stegen,keith}).

Our work shows that the allometric theories are a powerful tool for predicting forest features but that caution should be applied in using these theories as many processes, such as disturbance effects, can dramatically alter the observed scaling relationships. It should be noted that if the true upper bound on tree height can be predicted then disturbance processes can be calculated from the theory presented here. For example, NPP will be constant across all size classes and so the NPP calculated for the largest tree should represent the observed value even in a disturbed forest providing it fully utilises all existing resources. Similarly for biomass density one can use the predicted upper bound to calculate $n_{k}$ and then integrate equation \ref{masstot} up to the observed maximum size to get the true biomass density of the disturbed forest. In this later case this will lead to lower biomass density in recently disturbed forests because young trees have a higher metabolic rate per unit mass ($\alpha_{1}<1$). These calculations connect the steady-state work presented here to future efforts to quantify disturbance processes and highlight the importance of predicting the upper bound on tree height which is the subject of ongoing efforts (e.g. \cite{kempes,shi,ni}). Furthermore, an underlying assumption of this model is that trees seek to be as tall as possible \cite{king}, however this is dependent on the types of species that live in a given environment and the true upper bound on height may not be achievable if climate has changed recently and species have been unable to migrate (e.g. \cite{malcolm, loarie}). This may be an additional source of error between the predicted upper bound on tree height and observed values of NPP or biomass density.

It should be noted that most of the perspectives used within this analysis are based on fundamental physical constraints or basic biological processes such as mechanical stability, space filling, resource competition, and mortality, and these results represent numerous detailed derivations and predictions from a large number of researchers. The final product, however, is conceptually simple in that total forest net primary productivity and biomass density scale with the stem radius of the largest tree according to exponents of $2/3$ and $4/3$. Furthermore, our analysis here shows how various allometric perspectives can be combined to predict forest structure from local resources. The key link is that many forest features can be derived from maximum tree size which in turn is constrained by meteorological conditions. Moving forward these results provide a reduced perspective on key Earth-system processes such as biomass production rates and can be used to forecast both ecological response to changing climate and also the biogenic climate feedback. In addition, it will be important to integrate these perspectives with the rich history and current efforts of ecological and climate models that are able to handle detailed species-level differences, real-time forest dynamics and disturbances, the complex processes of soil dynamics, and the interconnection between forest processes and transient meteorology \cite{fisher,chapin,bonan2008,medvigy,bonan,running-2,colin,moorcroft-2,purves-2}.

\begin{methods}
Field-measured stem radius was obtained from the Forest Inventory and Analysis (FIA) data \cite{gray} spanning years from 2003 to 2007 and representing over two million trees. The aboveground biomass density map (circa 2003) was derived from the USDA Forest Service \cite{blackard} while we obtained the NPP estimates from a product (MOD17A3) of the Moderate Resolution Imaging Spectroradiometer (MODIS) averaged from 2001 to 2005 \cite{zhao}. For the forest stand ages, the North American Carbon Program (NACP) data were used (circa 2006) \cite{pan}. This MODIS product measures the surface absorption of Photosynthetically Active Radiation (APAR) and converts the APAR to the NPP based on the radiation use efficiency parameter, vegetation type, and climatic conditions (e.g. \cite{zhao-2}). Although the data has been validated with Eddy flux data, ecosystem model simulations, and atmospheric CO$_{2}$ measurements \cite{zhao-2}, it should be noted that such spaceborne information may have errors resulting from aerosol/cloud contamination and saturation in dense vegetation (e.g., \cite{samanta}). 

All gridded data were resampled at a 1-km spatial resolution, where, for stem radius we used a 3 km by 3 km moving window to search for the FIA maximum in order to correct for the random ``fuzzing'' that has been intentionally applied by the Forest Service and introduces errors up to 1.6 km \cite{gray}. Climate data used in this study were derived from the Parameter-elevation Regressions on Independent Slopes Model (PRISM) \cite{daly}. The PRISM provided long-term mean annul total precipitation at the 800 m spatial resolution (climatological averages of 1971-2000 \cite{prism}). The ecological region map \cite{ecoregion} was implemented to classify the data within each age bin ({\bf Table 1}). We used an eco-regional unit, i.e. province (n = 36). According to Cleland et al. \cite{cleland}, each boundary embraces distinctive biotic and environmental factors governing ecosystem structure and function: provinces represent climatic subzones in relation to geographical location and vegetation type.

For each of the observed scaling relationships we found exponents by applying both nonlinear least squares fitting of the power law to the raw data and linear fits to the logarithmically transformed data. The results from the two methods were similar ({\bf Figure} \ref{scaling-exponents} and {\bf Figure} \ref{extended-scaling}). For predictions of $r_{max}$, and subsequently $NPP$, from precipitation we solved for the intersection of basal metabolism ($Q_{0}$) with available water ($Q_{p}$) as described in Kempes et al. \cite{kempes}. The key parameter here is $\gamma$, the water absorption efficiency, reflecting the local soil and terrain properties. The hypothetical $\gamma$ of $1/3$ from Kempes et al. 2011 was updated using the topographic wetness index (TWI) \cite{beven}, which takes into account both water-flow direction and accumulation. For the continental United States, we calculated and normalized the TWI with that of a flat area (relative wetness index: rTWI$\equiv1$ for a flat area). We found that the mean of the rTWI over the continental United States is $0.21$, which we use for $\gamma$.

\end{methods}

\section*{Publication Note}
A previous version of this manuscript was originally submitted to a peer reviewed journal on March 20, 2015, and following referee comments this current version represents the revisions that have been recently submitted to the same journal on May 30, 2015.

\bibliography{tree}

\begin{addendum}
 \item The authors thank Suzanne Kern for comments on the manuscript. C.P.K thanks the Santa Fe Institute for support. S.C. was supported by the Fulbright Program for graduate studies and the NASA Earth and Space Science Fellowship Program (Grant NNX13AP55H). GBW would like to thank the John Templeton Foundation (grant no. 15705) and the Eugene and Clare Thaw Charitable Trust for their generous support.
 \item[Competing Interests] The authors declare that they have no
competing financial interests.
 \item[Correspondence] Correspondence and requests for materials
should be addressed to C.P.K.~(email: ckempes@gmail.com).
\end{addendum}

\clearpage
\newpage

\renewcommand{\figurename}{{\bf Figure}}
\renewcommand{\thefigure}{{\bf \arabic{figure}}}


\pagestyle{empty}

\begin{figure}
\centering
\subfigure{
\includegraphics[width=.48\textwidth]{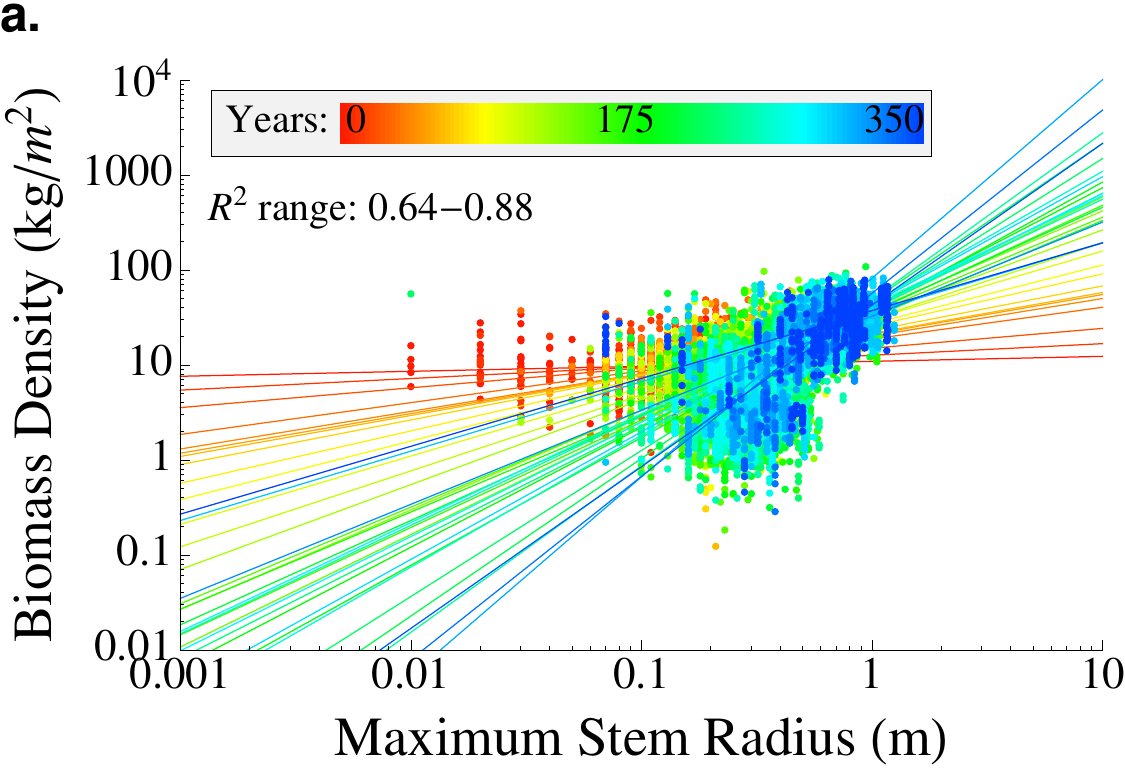}
\label{biomass-scaling}
}
\setcounter{subfigure}{3}
\subfigure{
\includegraphics[width=.48\textwidth]{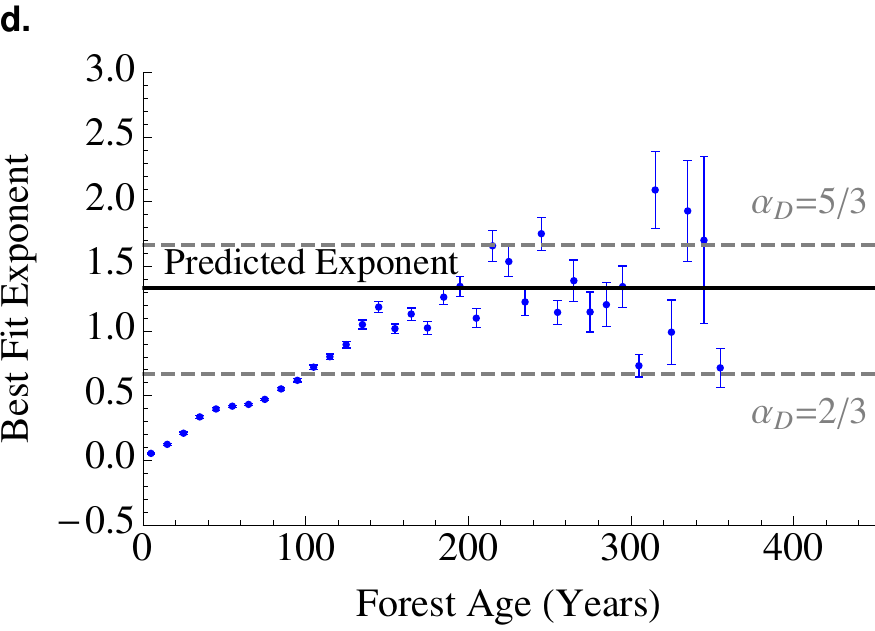}
\label{biomass-age}
}
\setcounter{subfigure}{1}
\subfigure{
\includegraphics[width=.48\textwidth]{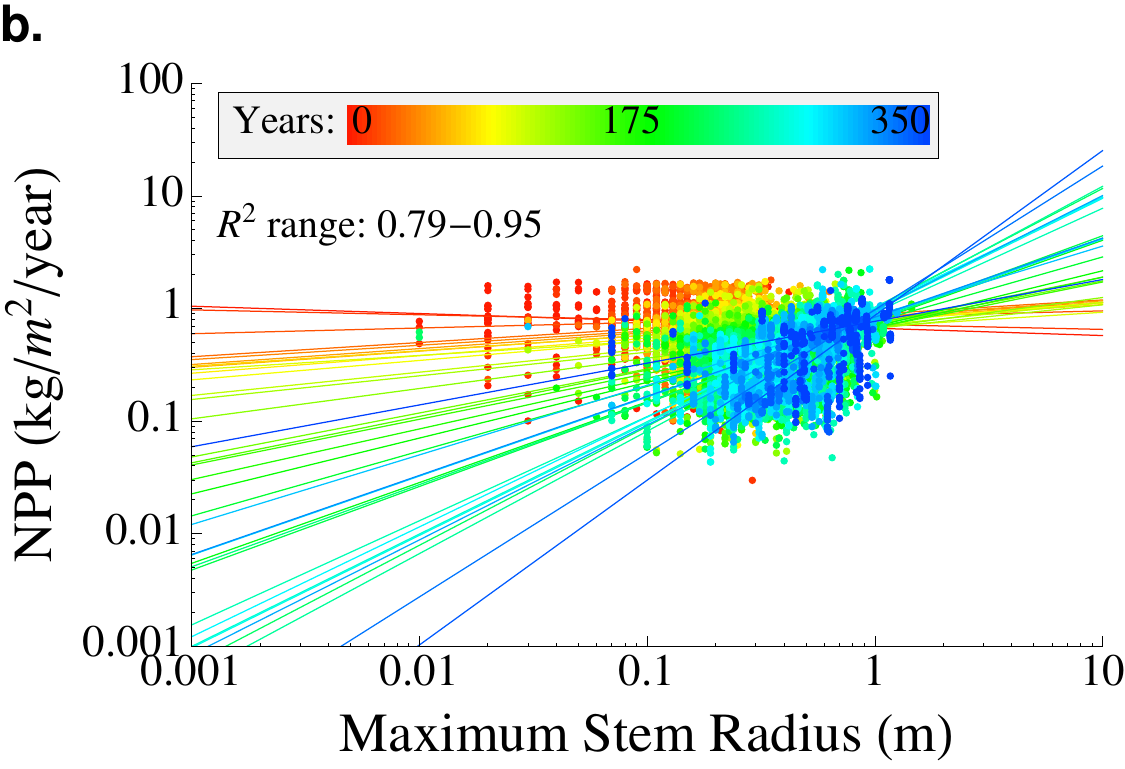}
\label{npp-scaling}
}
\setcounter{subfigure}{4}
\subfigure{
\includegraphics[width=.48\textwidth]{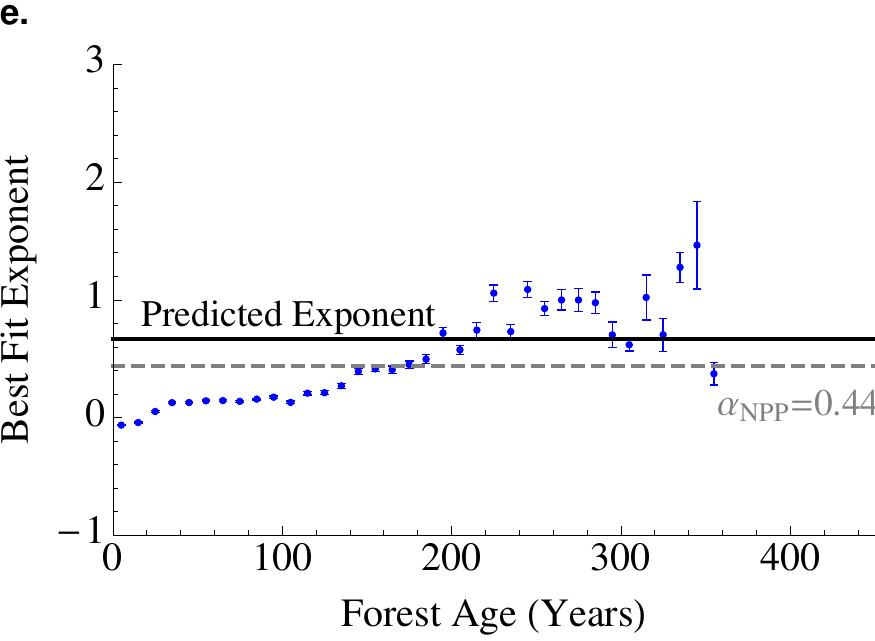}
\label{npp-age}
}
\setcounter{subfigure}{2}
\subfigure{
\includegraphics[width=.48\textwidth]{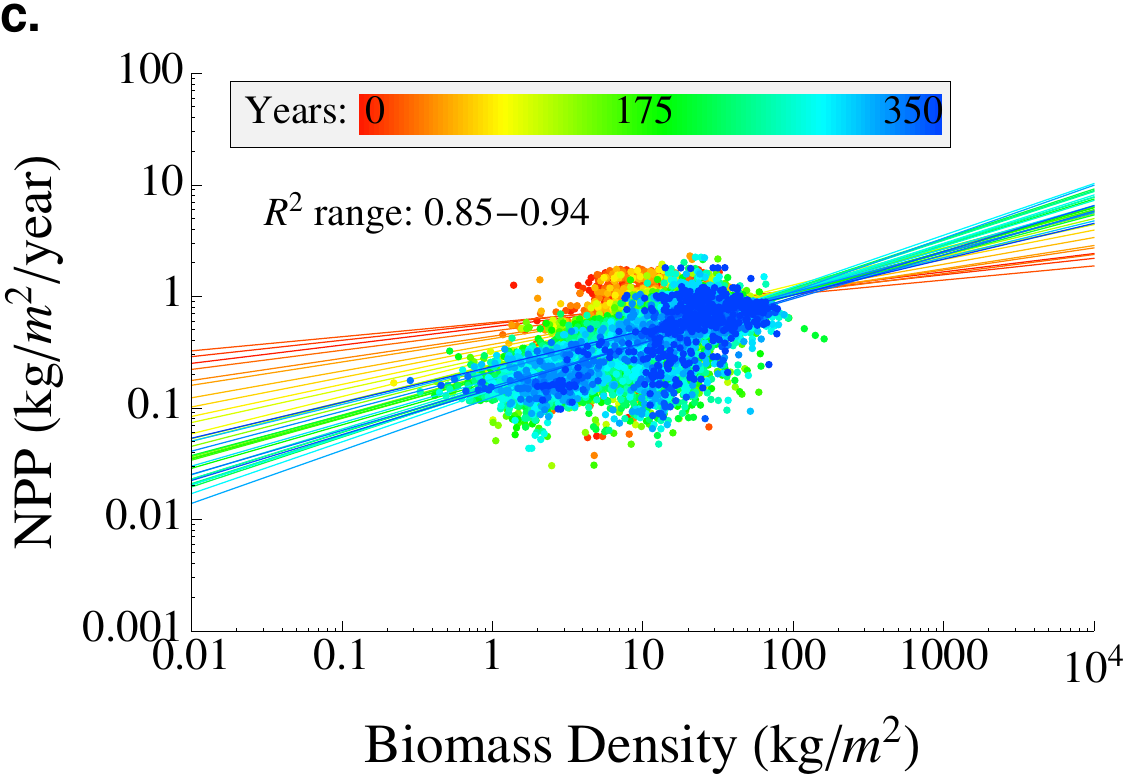}
\label{npp-biomass-scaling}
}
\setcounter{subfigure}{5}
\subfigure{
\includegraphics[width=.48\textwidth]{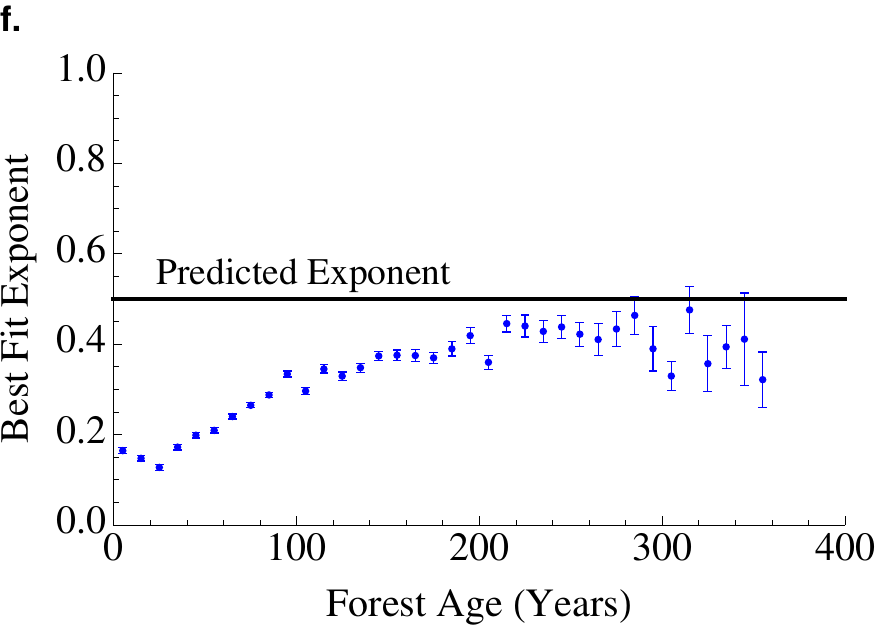}
\label{npp-biomass-age}
}
\caption{\linespread{1.0}\selectfont{} Nonlinear fits for the scaling of {\bf\subref{biomass-scaling}} biomass density and {\bf \subref{npp-scaling}} net primary productivity based on the maximum tree size as represented by trunk radius. {\bf \subref{npp-biomass-scaling}} The relationship between biomass density and net primary productivity. {\bf \subref{biomass-age}-\subref{npp-biomass-age}} The corresponding best fit exponents as a function of forest age class compared with the theoretical predictions. The data were analyzed within 10-year forest age bins for forests ranging from 10 to 350 years old. For the scaling relationships 500 randomly selected data points are shown for visual clarity. However, in fitting the scaling relationships we used all available data (909,524 grid cells) and each age bin contained an average of 25,265 data points, ranging between 121 for the 340-350 year age bin to 102,826 for the 60-70 year age bin, all in the continental United States (data from\cite{gray,pan,blackard,zhao}). The range of $R^{2}$ values for power-law fits from all age classes are given and demonstrate a strong correlation with a minimum of $R^{2}=.64$ across all data. The grey dashed lines represent alternative values for the exponents as discussed in the text.}
\label{scaling-exponents}
\end{figure}


\begin{figure}
\centering
\subfigure{
\includegraphics[width=.52\textwidth]{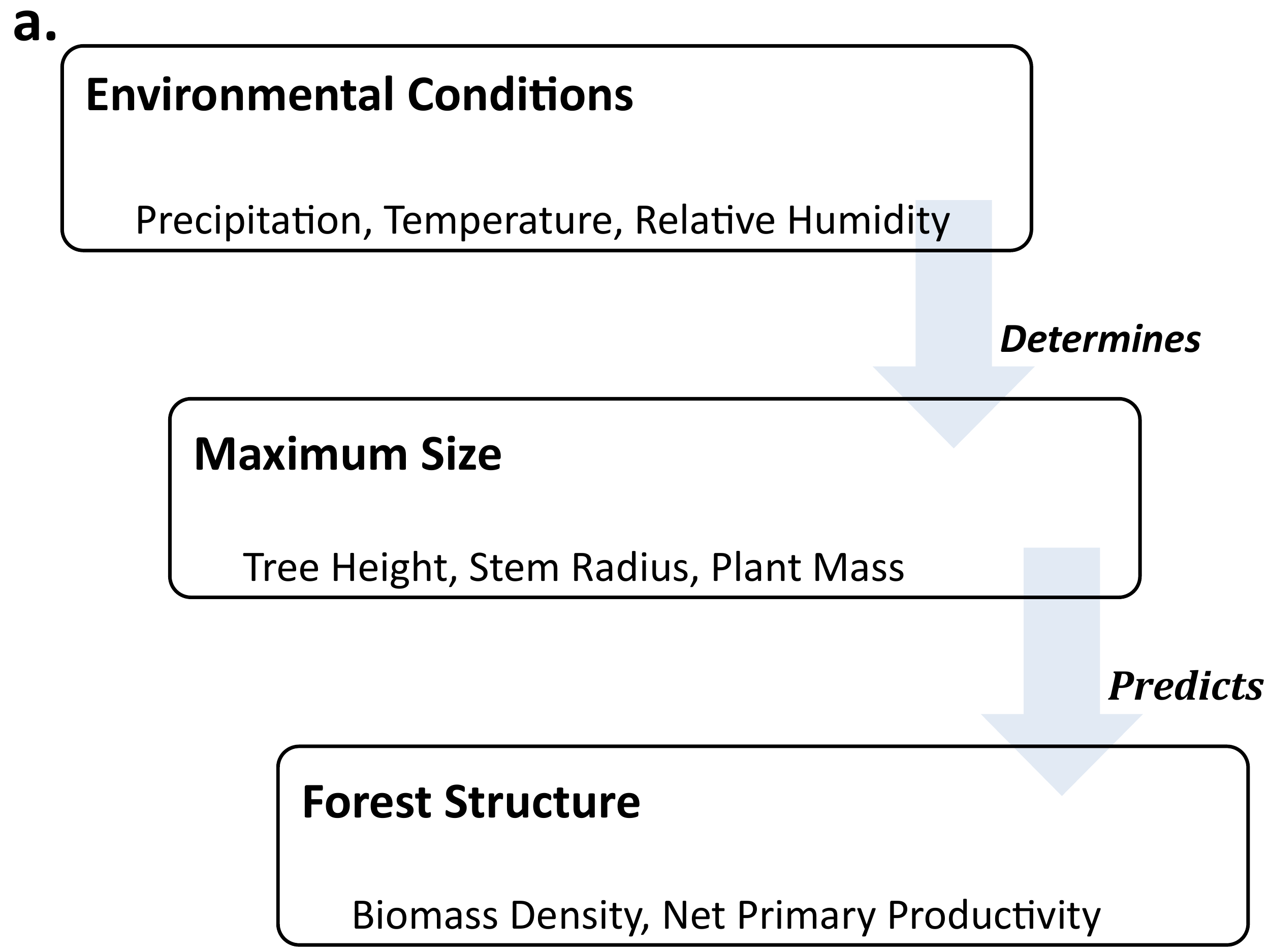}
\label{framework-schematic}
}
\subfigure{
\includegraphics[width=.48\textwidth]{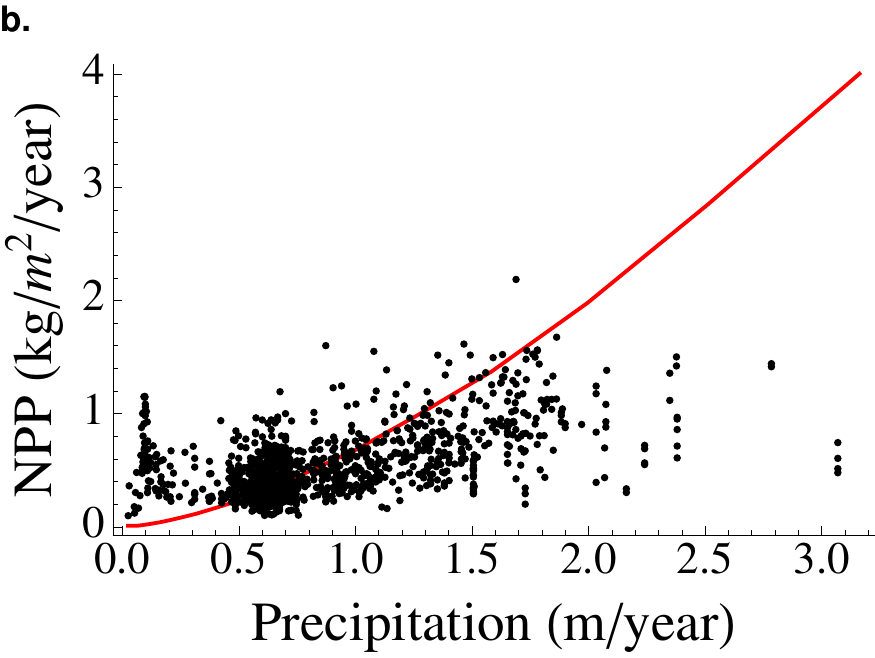}
\label{precip-scaling}
}
\subfigure{
\includegraphics[width=.48\textwidth]{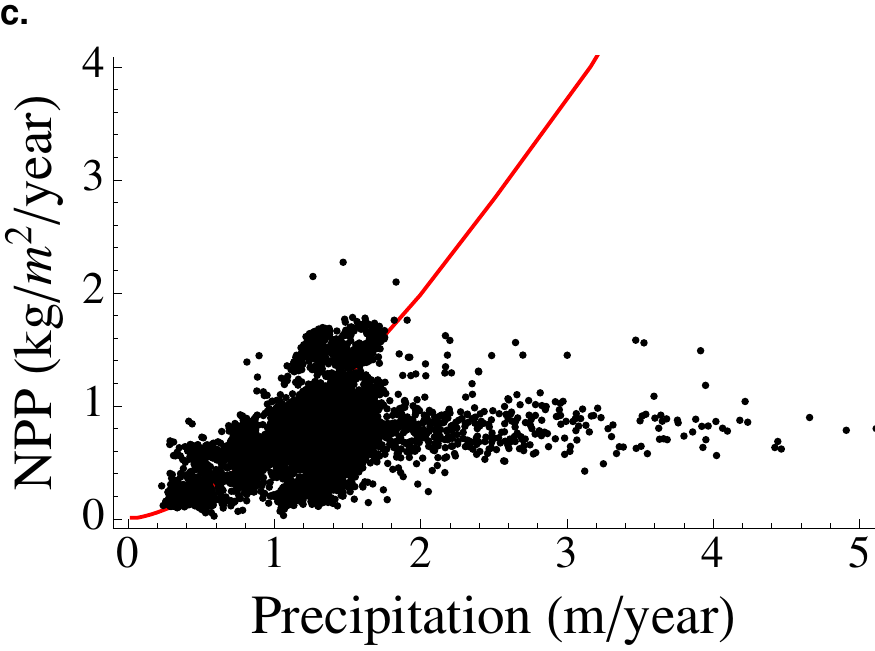}
\label{our-precip-scaling}
}
\caption{\linespread{1.0}\selectfont{} {\bf \subref{framework-schematic}} A schematic of the framework for predicting bulk forest features from local resources. Predictions from this framework (red line) for the dependence of NPP on annual precipitation for {\bf \subref{precip-scaling}} the data from Michaletz et al. \cite{michaletz} and {\bf \subref{our-precip-scaling}} our own analysis for the United States (10,000 randomly selected points are shown for visual clarity). Maximum tree size is predicted using the framework presented in Kempes et al. \cite{kempes} and converted to NPP using the results from this paper. The prediction matches the data well with R$^{2}=0.82$ for {\bf \subref{precip-scaling}} and R$^{2}=0.46$ for {\bf \subref{our-precip-scaling}}.}
\label{environmental-predictions}
\end{figure}


\begin{figure}
\centering
\subfigure{
\includegraphics[width=.48\textwidth]{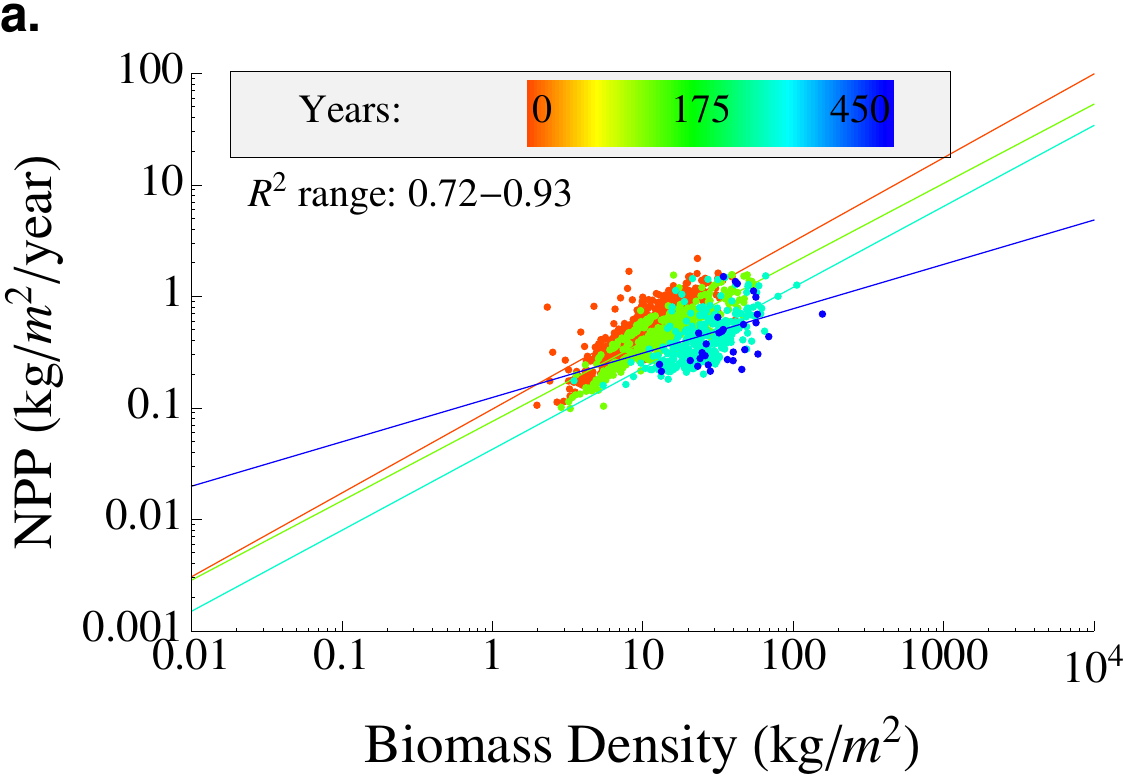}
\label{enquist-npp-biomass}
}
\subfigure{
\includegraphics[width=.48\textwidth]{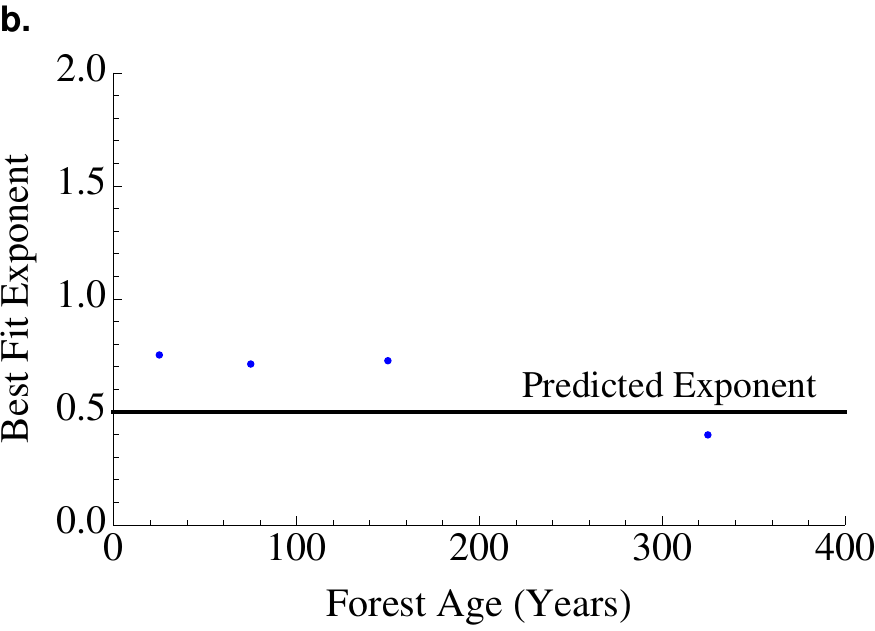}
\label{enquist-npp-biomass-age}
}
\caption{\linespread{1.0}\selectfont{} The scaling relationships and exponents for biomass and net primary productivity for the data presented in Michaletz et al. \cite{michaletz}. The scaling exponents approach the predicted value of $.5$ for mature forests.}
\label{enquist-npp}
\end{figure}

\begin{figure}
\centering
\subfigure{
\includegraphics[width=.48\textwidth]{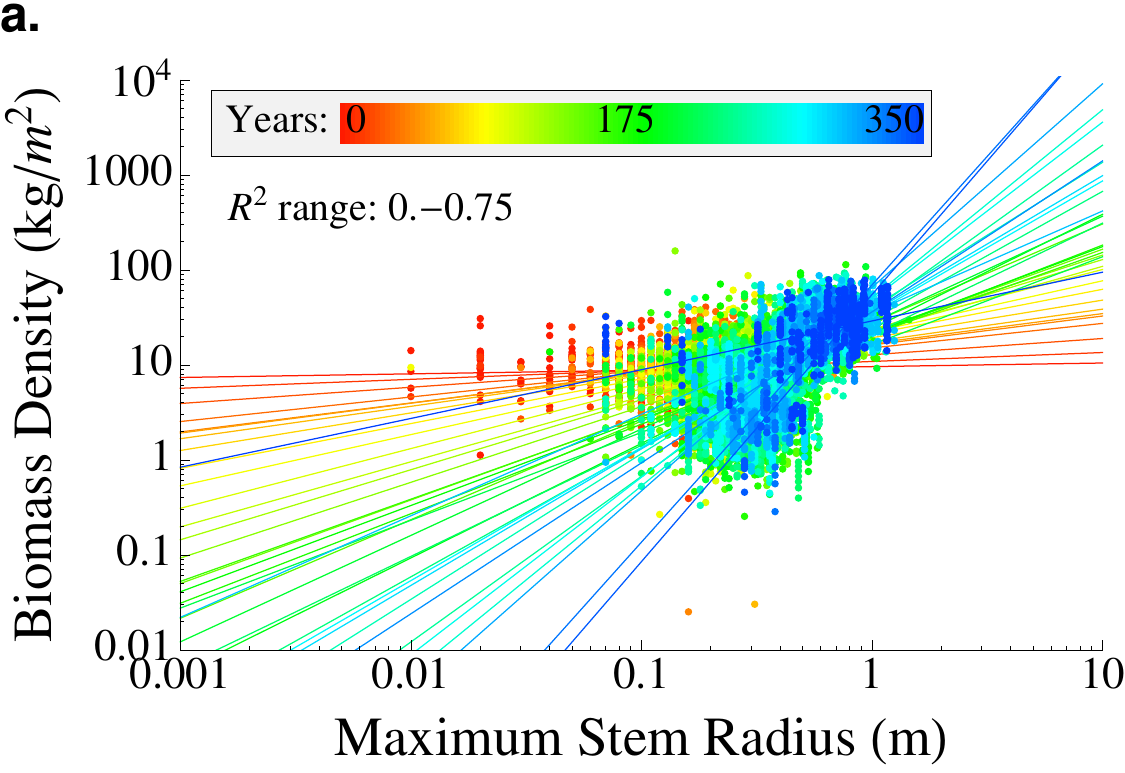}
\label{biomass-scaling-2}
}
\setcounter{subfigure}{3}
\subfigure{
\includegraphics[width=.48\textwidth]{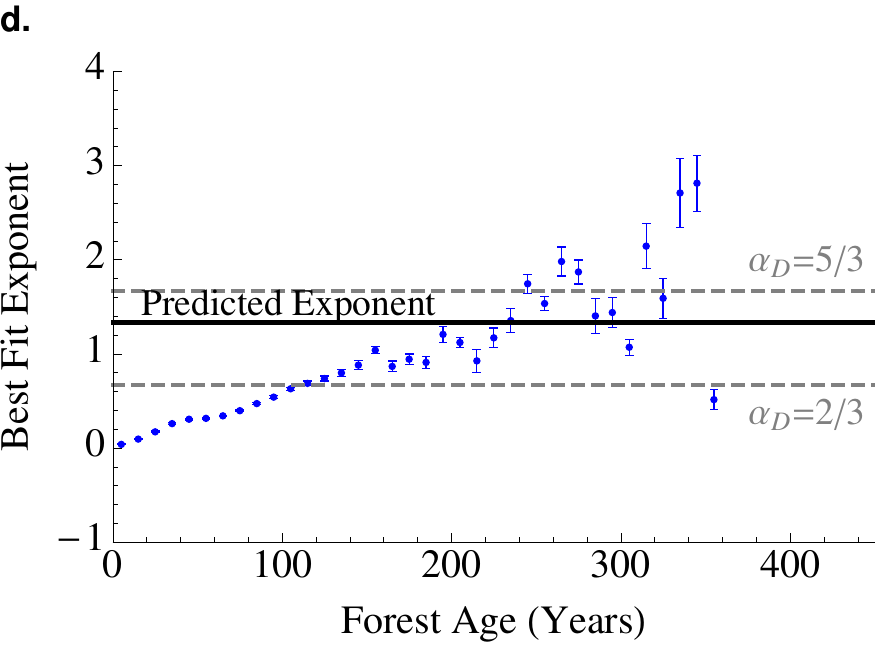}
\label{biomass-age-2}
}
\setcounter{subfigure}{1}
\subfigure{
\includegraphics[width=.48\textwidth]{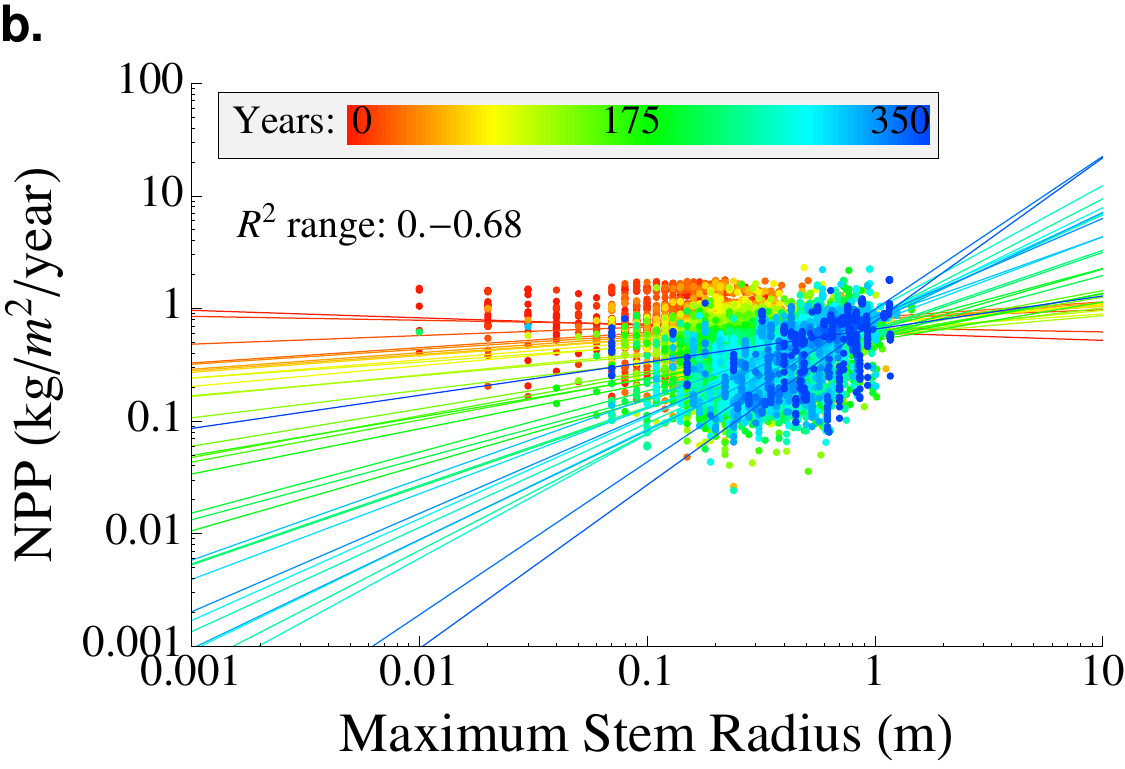}
\label{npp-scaling-2}
}
\setcounter{subfigure}{4}
\subfigure{
\includegraphics[width=.48\textwidth]{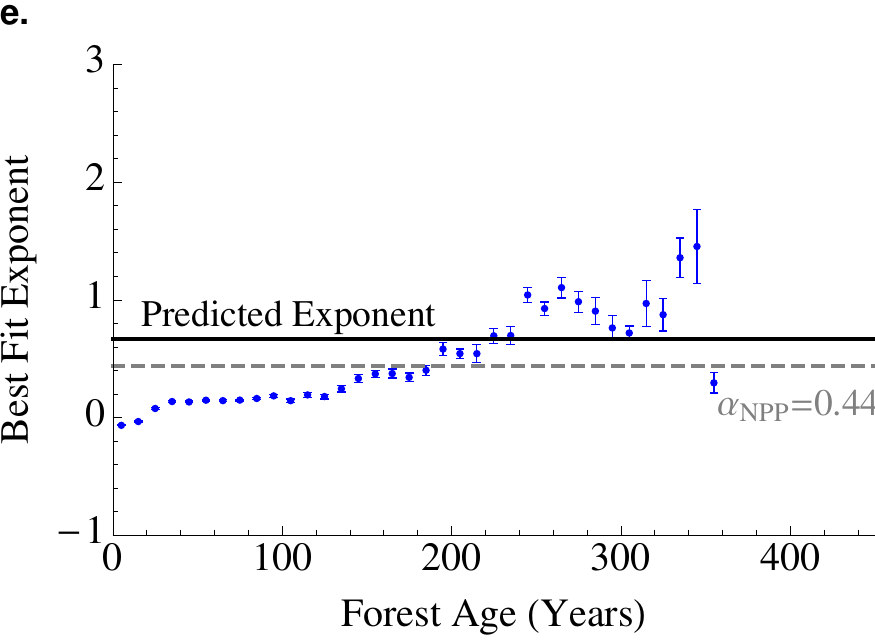}
\label{npp-age-2}
}
\setcounter{subfigure}{2}
\subfigure{
\includegraphics[width=.48\textwidth]{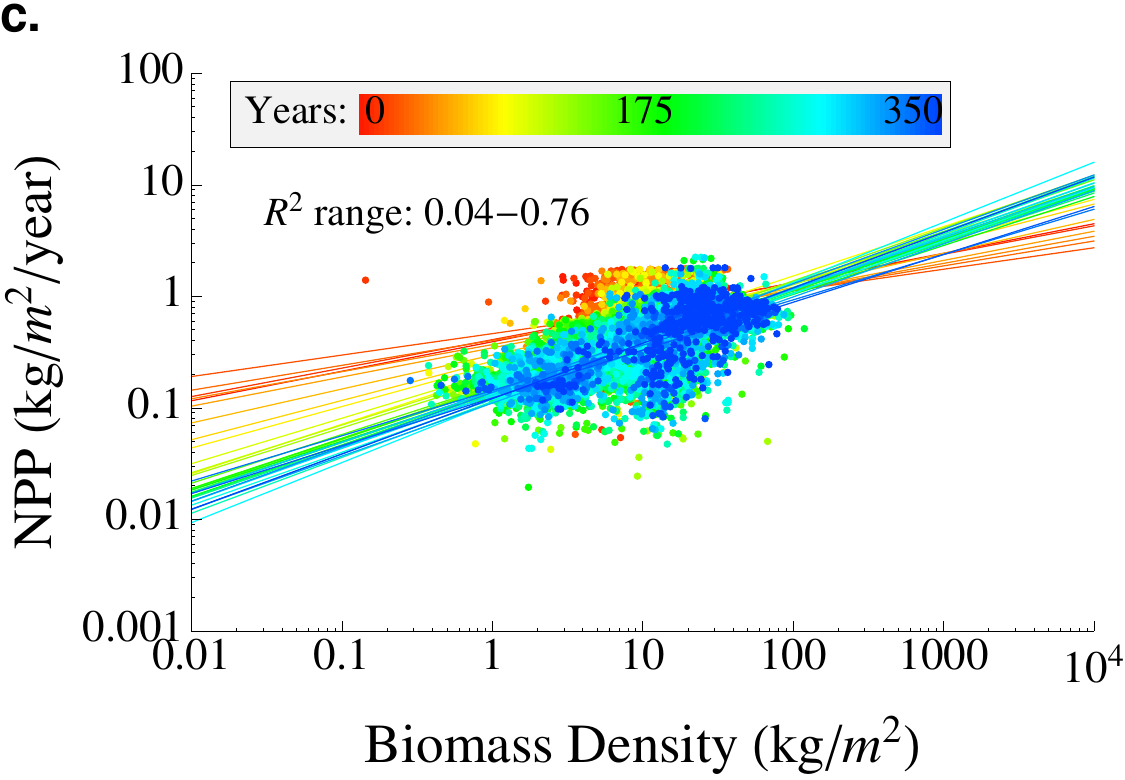}
\label{npp-biomass-scaling-2}
}
\setcounter{subfigure}{5}
\subfigure{
\includegraphics[width=.48\textwidth]{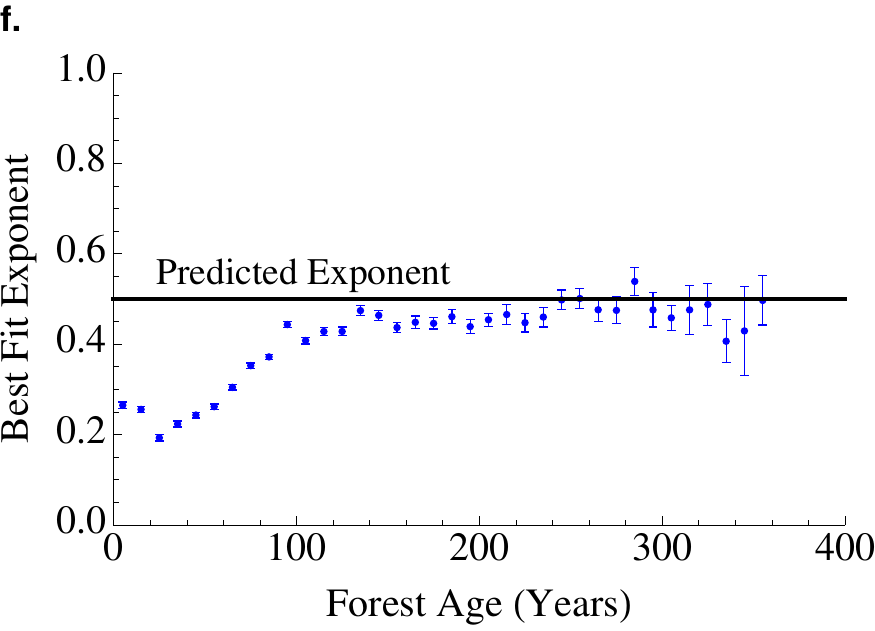}
\label{npp-biomass-age-2}
}
\caption{\linespread{1.0}\selectfont{} {\it Linear fits of logarithmically transformed data} for the scaling of {\bf \subref{biomass-scaling-2}} biomass density and {\bf \subref{npp-scaling-2}} net primary productivity (NPP) based on the maximum tree size as represented by trunk radius. {\bf \subref{npp-biomass-scaling-2}} The relationship between biomass density and net primary productivity. {\bf \subref{biomass-age-2}-\subref{npp-biomass-age-2}} The corresponding best fit exponents as a function of forest age class compared with the  theoretical predictions. The data were analyzed within 10 year forest age bins for forests ranging from 10 to 350 years old. For the scaling relationships 500 randomly selected data points are shown for visual clarity. However, in fitting the scaling relationships we used all available data (909,524 grid cells) and each age bin contained an average of 25,264 data points, ranging between 121 for the 340-350 year age bin to 102,826 for the 60-70 year age bin, all from the continental United States (data from \cite{gray,pan,blackard,zhao}). The grey dashed lines represent alternative values for the exponents as discussed in the text.}
\label{extended-scaling}
\end{figure}

\begin{table}
\tiny
\centering
\caption{Statistics for allometric fits within each age bin.}
\medskip
\begin{tabular}{p{.75cm}|p{.75cm}|p{1.cm}|p{1.cm}|p{1.cm}|p{.5cm}|p{1.cm}|p{.5cm}|p{1.cm}|p{.5cm}}
\hline
Age Bin (Years) & Number of Plots & Number of Ecoregions & Spans Arid to Temperate Forests & Biomass vs. $r_{max}$ Exponent & $R^{2}$& NPP vs. $r_{max}$ Exponent & $R^{2}$ & NPP vs. Biomass Exponent & $R^{2}$ \\
\hline
0-10 & 71983 & 34 & Yes & 0.05 & 0.81 & -0.07 & 0.86 & 0.16 & 0.86\\
10-20 & 97112 & 35 & Yes & 0.12 & 0.81 & -0.04 & 0.86 & 0.15 & 0.86\\
20-30 & 76794 & 35 & Yes & 0.21 & 0.81 & 0.05 & 0.86 & 0.13 & 0.86\\
30-40 & 76277 & 35 & Yes & 0.33 & 0.82 & 0.13 & 0.87 & 0.17 & 0.87\\
40-50 & 86753 & 35 & Yes & 0.4 & 0.85 & 0.13 & 0.88 & 0.2 & 0.88\\
50-60 & 99391 & 35 & Yes & 0.42 & 0.86 & 0.14 & 0.89 & 0.21 & 0.89\\
60-70 & 102826 & 35 & Yes & 0.43 & 0.88 & 0.14 & 0.89 & 0.24 & 0.89\\
70-80 & 89694 & 35 & Yes & 0.47 & 0.87 & 0.14 & 0.9 & 0.26 & 0.91\\
80-90 & 65645 & 35 & Yes & 0.55 & 0.86 & 0.16 & 0.9 & 0.29 & 0.91\\
90-100 & 38584 & 35 & Yes & 0.62 & 0.84 & 0.17 & 0.89 & 0.33 & 0.92\\
100-110 & 25920 & 33 & Yes & 0.72 & 0.8 & 0.13 & 0.87 & 0.3 & 0.9\\
110-120 & 14991 & 31 & Yes & 0.8 & 0.78 & 0.21 & 0.85 & 0.34 & 0.89\\
120-130 & 11877 & 30 & Yes & 0.89 & 0.75 & 0.21 & 0.85 & 0.33 & 0.9\\
130-140 & 8442 & 27 & Yes & 1.05 & 0.71 & 0.27 & 0.84 & 0.35 & 0.9\\
140-150 & 6082 & 24 & Yes & 1.18 & 0.73 & 0.39 & 0.84 & 0.37 & 0.9\\
150-160 & 7287 & 26 & Yes & 1.02 & 0.74 & 0.41 & 0.82 & 0.38 & 0.88\\
160-170 & 4430 & 22 & Yes & 1.13 & 0.71 & 0.41 & 0.83 & 0.37 & 0.89\\
170-180 & 4406 & 23 & Yes & 1.02 & 0.64 & 0.45 & 0.8 & 0.37 & 0.88\\
180-190 & 3479 & 17 & Yes & 1.26 & 0.71 & 0.5 & 0.8 & 0.39 & 0.87\\
190-200 & 2550 & 19 & Yes & 1.34 & 0.69 & 0.72 & 0.79 & 0.42 & 0.86\\
200-210 & 3601 & 17 & Yes & 1.1 & 0.67 & 0.57 & 0.82 & 0.36 & 0.88\\
210-220 & 1382 & 13 & Yes & 1.66 & 0.66 & 0.74 & 0.81 & 0.45 & 0.91\\
220-230 & 1467 & 14 & Yes & 1.54 & 0.68 & 1.06 & 0.8 & 0.44 & 0.85\\
230-240 & 1063 & 12 & Yes & 1.22 & 0.69 & 0.73 & 0.82 & 0.43 & 0.89\\
240-250 & 1143 & 13 & Yes & 1.75 & 0.76 & 1.09 & 0.86 & 0.44 & 0.88\\
250-260 & 1878 & 14 & Yes & 1.14 & 0.77 & 0.93 & 0.85 & 0.42 & 0.86\\
260-270 & 535 & 11 & Yes & 1.39 & 0.76 & 1. & 0.88 & 0.41 & 0.89\\
270-280 & 663 & 12 & Yes & 1.15 & 0.73 & 1. & 0.84 & 0.43 & 0.87\\
280-290 & 382 & 9 & Yes & 1.2 & 0.73 & 0.98 & 0.87 & 0.46 & 0.9\\
290-300 & 422 & 11 & Yes & 1.34 & 0.81 & 0.7 & 0.82 & 0.39 & 0.86\\
300-310 & 1284 & 12 & Yes & 0.73 & 0.8 & 0.62 & 0.89 & 0.33 & 0.9\\
310-320 & 247 & 8 & Yes & 2.09 & 0.79 & 1.02 & 0.79 & 0.48 & 0.89\\
320-330 & 213 & 8 & Yes & 0.99 & 0.79 & 0.7 & 0.88 & 0.36 & 0.9\\
330-340 & 124 & 7 & Yes & 1.93 & 0.81 & 1.28 & 0.95 & 0.39 & 0.94\\
340-350 & 121 & 5 & Yes & 1.7 & 0.79 & 1.46 & 0.89 & 0.41 & 0.88\\
350-360 & 476 & 5 & Yes & 0.71 & 0.8 & 0.37 & 0.84 & 0.32 & 0.86\\
\hline
\end{tabular}
\end{table}

\end{document}